\documentclass[12pt]{article}
\usepackage[english]{babel}
\usepackage[utf8]{inputenc}
\usepackage[T1]{fontenc}

\usepackage{amsmath}
\usepackage{amsfonts}
\usepackage{url}
\usepackage[dvips]{graphicx}
\usepackage{calc}
\usepackage{subfigure}
\usepackage{multirow}

\def\mL{\mathcal{L}}

\def\mH{\mathcal{H}}

\def\bx{\mathbf{x}}
\newcommand{\pb}[1]{\left\{#1\right\}}

\def\tX{\tilde{X}}

\def\mG{\mathcal{G}}
\def\by{\mathbf{y}}

\def\bT{\mathbf{T}}

\begin{document}
	\begin{titlepage}

	\vskip 0.4 cm
	
	\begin{center}
		{\Large{ \bf Note About Canonical Formalism for Gravity with Dynamical Determinant
		of Metric}}
		
		\vspace{1em}  
		
		\vspace{1em} J. Kluso\v{n} 			
		\footnote{Email addresses:
			klu@physics.muni.cz (J.
			Kluso\v{n}) }\\
		\vspace{1em}
		\textit{Department of Theoretical Physics and
			Astrophysics, Faculty of Science,\\
			Masaryk University, Kotl\'a\v{r}sk\'a 2, 611 37, Brno, Czech Republic}
		
		\vskip 0.8cm
		
		%
		%
		%
		%
		%
		%
		
		\vskip 0.8cm
		
	\end{center}

	\begin{abstract}
In this short note we perform canonical analysis of  the theory invariant 
under restricted diffeomorphism so that the action contains kinetic term for determinant of metric. We find corresponding Hamiltonian and determine structure of constraints.	
		
	\end{abstract}
	
	\bigskip
	
\end{titlepage}

\newpage

\section{Introduction and Summary}\label{first}
Einstein's theory of gravity is one  of the  most successful physical theories. Characteristic property of this theory that the equations of motion can be 
derived from the action that is invariant under general coordinate transformations. In fact,  the Principe of General Covariance is generally accepted as the basic tool for constructions of field theories interacting with gravity. 

On the other hand it was shown long time ago in \cite{Buchmuller:1988wx} that even 
smaller symmetric group is sufficient for determination of Einstein equations of motion. Famous example of such a theory is unimodular gravity  
\cite{Henneaux:1989zc,Kuchar:1991xd,Unruh:1988in} where allowed coordinate transformations preserve volume element.  Interesting property of unimodular gravity  is that constant term that appears in the action does not correspond to the cosmological constant. In fact, cosmological constant arises as integration constant when we impose  condition  that matter stress energy tensor should be preserved, for recent review of unimodular gravity, see for example 
\cite{Jirousek:2023gzr,Alvarez:2023utn,Carballo-Rubio:2022ofy}, for same recent works, see \cite{Garay:2023nco,Kehagias:2022mik,Tiwari:2022ctc,Alonso-Serrano:2021uok,Alonso-Serrano:2020pcz,Nojiri:2015sfd}. Further, due to the presence of the unimodular constraint
canonical formulation  of unimodular gravity is also non-trivial \cite{Karataeva:2022mll,Bufalo:2017tms,Bufalo:2015wda,Kluson:2014esa}.

However it is important to stress that relaxing the full diffeomorphism invariance to the 
restricted ones allows more general theories than the unimodular ones. Such an example of this theory was presented in \cite{Alvarez:2006uu} where the general theories invariant under transverse diffeomorphism were analysed. It turns out that  these theories allow kinetic term for the determinant of metric. Important example of such a theory is Weyl transverse gravity \cite{Oda:2016vui,Oda:2016psn} which is invariant under transverse diffeomorphism and Weyl rescaling of metric as well. Gauge fixing of this Weyl symmetry we obtain unimodular gravity that makes connection between these theories transparent. Further, Weyl transverse gravity has the same classical solutions as general relativity while it can solve some of the problems related to cosmological constant as was discussed recently in \cite{Alonso-Serrano:2022pif,Alonso-Serrano:2022rzj}.

Let us discuss in more details the fact that kinetic term for the determinant of  metric
is allowed in the  theory invariant under transverse diffeomorphism. If we express this term using $D+1$ decomposition of the metric
\cite{Arnowitt:1962hi} we immediately find that there is a kinetic term for the lapse function which is in sharp contrast with General Relativity case when such a term is missing. 
The presence of kinetic term for lapse in theory invariant under restricted diffeomorphism 
has non-trivial consequence on the canonical structure of this theory since momentum conjugate to lapse is no longer primary constraint. In this paper we restrict ourselves to the specific model of gravity invariant under restricted diffeomorphism leaving the most general one, including Weyl transverse gravity, for future work. More precisely, in the model discussed in this paper we can ignore total derivative terms that appear in $D+1$ decomposition of Ricci scalar 
\cite{Gourgoulhon:2007ue}. Then we determine corresponding Hamiltonian when we express time derivative of determinant of  metric with the help of variables used in $D+1$ decomposition. We immediately identify primary constraints which are momenta conjugate to shift functions since, despite of the presence of the kinetic term for determinant, the time derivative of shift functions are still missing. As the next step we study time evolution of the primary constraints and we find that they are preserved when we introduce secondary constraints which will be interpreted as generators of spatial diffeomorphism with one subtle difference with respect to the full diffeomorphism invariant case. Due to the presence of kinetic term for lapse function these secondary constraints have terms proportional to the momentum conjugate to lapse so that there is non-trivial Poisson bracket between smeared form of spatial diffeomorphism constraints and lapse function.  This result has crucial consequence when  we impose requirement of the preservation of  secondary constraints and it turns out that they will be preserved when we introduce new, Hamiltonian constraint which now behaves as scalar rather than scalar density.  In other words Hamiltonian constraint emerges in this specific model as well  and the last step will be analysis  of the preservation of this constraint during time evolution.  We find
that the Hamiltonian constraint is the first class constraint  and we also find that the Poisson brackets between spatial diffeomorphism constraints and Hamiltonian constraint have the same form as in ordinary general relativity theory.

Let us outline our results and suggest  possible extension of this work. Main goal of this paper is to perform canonical analysis of specific model of gravity where the action contains kinetic term for determinant of metric. This model is also known as theory invariant under restricted diffeomorphism. We found its canonical form and we also determined structure of constraints. The main result derived here is that the Poisson brackets between these constraints have the same form as in case of General Relativity. Then the structure of constraints determine number of physical degrees of freedom. We found that there is one additional mode with respect to General Relativity case due to the fact that  lapse function is dynamical. On the other hand this mode could be eliminated when the theory invariant under restricted diffeomorphism possesses additional gauge symmetry. Such an important example is Weyl transverse gravity which was shown to be equivalent to General Relativity theory in the sense that solutions of their equations of motion are the same as in General Relativity. Then it would be extremely interesting to study canonical structure of Weyl gravity and compare it corresponding analysis performed in case of General Relativity. This work is currently in the progress. 

This paper is organized as follows. In the next section (\ref{second}) we introduce action invariant under restricted diffeomorphism and find corresponding Hamiltonian. Then in section (\ref{third}) we study constraint's structure of this theory.

\section{Action with Dynamical determinant of Metric}\label{second}
In this section we introduce an action where the determinant of the metric is dynamical. 
In fact, such a theory corresponds to modified theories of gravity with restricted coordinate
invariance
\cite{Buchmuller:1988wx}. The simplest action with dynamical determinant of metric has the form  \cite{Alvarez:2006uu}   
\begin{eqnarray}\label{Sdet}
&&	S=\int d^{n}x\mL \ , \quad 
	\mL=\frac{1}{\kappa} \sqrt{-g}[R(g_{\mu\nu})+\frac{G(g)}{g^2}\partial_\mu g^{\mu\nu}
	\partial_\nu g] \ ,  \nonumber \\
&&	 g\equiv \det g \ , \quad \kappa=16\pi G \ , 
\end{eqnarray}
where $n$ is number of space-time dimensions and where $G(g)$ is arbitrary function of 
determinant of metric.  Before we proceed to the canonical formulation of this theory it is instructive to briefly discuss symmetry of the action (\ref{Sdet}).  We start with infinitesimal form of  general diffeomorphism 
\begin{equation}
	x'^\mu=x^\mu+\xi^\mu(x) \  
\end{equation}
that implies inverse relation 
\begin{equation}\label{inversetr}
	x^\mu=x'^\mu-\xi^\mu(x)
	\approx x'^\mu-\xi^\mu(x')+\mathcal{O}(\xi^2) \ , 
\end{equation}
where $\mu,\nu,\dots=0,1,\dots,(n-1)$. It is well known that the metric transforms under general diffeomorphism $x'^\mu=x'^\mu(x^\rho)$ as
\begin{equation}
	g'_{\mu\nu}(x')=\frac{\partial x^\rho}{\partial x'^\mu}\frac{\partial x^\sigma}{\partial x'^\nu}
	g_{\rho\sigma}(x) \ 
\end{equation}
that for (\ref{inversetr}) has the for 
\begin{equation}\label{mettran}
g'_{\mu\nu}(x')=g'_{\mu\nu}(x)+x^\rho\partial_\rho g_{\mu\nu}(x)=
-\partial_\mu \xi^\rho g_{\rho\nu}(x)-g_{\mu\rho}\partial_\nu\xi^\rho(x) \ . 
\end{equation}
Let us  define variation of the metric under diffeomorphism transformations  as $\delta g_{\mu\nu}(x)\equiv g'_{\mu\nu}(x)-g_{\mu\nu}(x)$. Then  (\ref{mettran}) implies that $\delta g_{\mu\nu}$ is equal to 
\begin{eqnarray}\label{deltag}
\delta g_{\mu\nu}(x)=	-\partial_\rho g_{\mu\nu}(x)\xi^\rho(x)
	-g_{\mu \rho}(x)\partial_\nu \xi^\rho(x)-\partial_\mu \xi^\rho(x) g_{\rho\nu}(x) \  
\end{eqnarray}
and consequently variation of the determinant of metric is equal to
\begin{equation}\label{varmetric}
	\delta g(x)=-\xi^\rho\partial_\rho g(x)-2\partial_\rho \xi^\rho g(x)  \ , \quad g=\det g_{\mu\nu} \ .  
\end{equation}
The second term in the expression above is crucial since the action 
 (\ref{Sdet}) will be invariant under  diffeomorphism transformation on condition when the last term in (\ref{varmetric}) vanishes. In other words the allowed diffeomorphism transformations have to obey the condition 
 \begin{equation}
 	\partial_\rho \xi^\rho=0 \ .
  \end{equation}

It is very instructive to perform canonical analysis of the theory defined by the action 
(\ref{Sdet}). Note that in principle we could consider more general form of the action when $\sqrt{-g}$ is replaced by arbitrary function $F(g)$. We leave discussion of this form of the action for future work.  

To proceed to the canonical formulation we use  $(n-1)+1$ splitting of space-time  that is the fundamental ingredient of the Hamiltonian
formalism of any theory of gravity \footnote{For recent review, see
	\cite{Gourgoulhon:2007ue}.}. We consider $n$ dimensional manifold
$\mathcal{M}$ with the coordinates $x^\mu \ , \mu=0,\dots,n-1$ and
where $x^\mu=(t,\bx) \ , \bx=(x^1,x^2,x^{n-1})$. We presume that this
space-time is endowed with the metric $g_{\mu\nu}(x^\rho)$
with signature $(-,+,\dots,+)$. Suppose that $ \mathcal{M}$ can be
foliated by a family of space-like surfaces $\Sigma_t$ defined by
$t=x^0$. Let $h_{ij}, i,j=1,2,3,n-1$ denotes the metric on $\Sigma_t$
with inverse $h^{ij}$ so that $h_{ij}h^{jk}= \delta_i^k$. We further
introduce the operator $\nabla_i$ that is covariant derivative
defined with the metric $h_{ij}$.
We also define  the lapse
function $N=1/\sqrt{-g^{00}}$ and the shift function
$N^i=-g^{0i}/g^{00}$. In terms of these variables we
write the components of the metric $g_{\mu\nu}$ as
\begin{eqnarray}\label{31metric}
	g_{00}=-N^2+N_i h^{ij}N_j \ , \quad g_{0i}=N_i \ , \quad
	g_{ij}=h_{ij} \ ,
	\nonumber \\
	g^{00}=-\frac{1}{N^2} \ , \quad g^{0i}=\frac{N^i}{N^2} \
	, \quad g^{ij}=h^{ij}-\frac{N^i N^j}{N^2} \ .
	\nonumber \\
\end{eqnarray}
and hence $g=-N^2\det h_{ij}\equiv -N^2h$. We further have following decomposition of scalar curvature $R$ in the form
\begin{eqnarray}\label{Rde}
	R=K^{ij}K_{ij}-K^2+r+\frac{2}{\sqrt{-g}}\partial_\mu[\sqrt{-g}n^\mu K]
	-\frac{2}{\sqrt{h}N}\partial_i[\sqrt{h}h^{ij}\partial_j N] \ , 
	\nonumber \\
	K_{ij}=\frac{1}{2N}(\partial_t h_{ij}-\nabla_i N_j-\nabla_j N_i) \ , 
	\quad n^0=\sqrt{-g^{00}} \ , \quad 
	n^i=-\frac{g^{0i}}{\sqrt{-g^{00}}} \ , \nonumber \\
\end{eqnarray}
and where $r$ is scalar curvature defined with $h_{ij}$.

The most interesting aspect of this theory is the kinetic term for determinant of the metric. Inserting the decomposition of the metric given in (\ref{31metric}) we obtain 
\begin{eqnarray}\label{kindet}
	\frac{1}{g^2}\partial_\mu g g^{\mu\nu}\partial_\nu g=
-\frac{1}{g^2}\frac{1}{N^2}(\partial_0 g-N^i\partial_i g)^2+
	\frac{1}{g^2}h^{ij}\partial_i g\partial_j g \ . \nonumber \\
\end{eqnarray}
To proceed further we perform following manipulation
\begin{eqnarray}\label{parnulaN}
&&	\frac{1}{Ng}(\partial_0 g-N^i\partial_i g)
	=2\frac{1}{N^2}(\partial_0 N-N^i\partial_i N)+\frac{1}{Nh}
	(\partial_0 h-N^i\partial_i h)= \nonumber \\
&&	=\frac{2}{N^2}(\partial_0N-N^i\partial_i N)+
	\frac{1}{N}(2NK_{ij}h^{ji}+2\nabla_i N_j h^{ji})-
	\frac{N^i}{N}\partial_i h_{kl}h^{kl}= 
	\nonumber \\
&&	=\frac{2}{N^2}(\partial_0 N-N^i\partial_i N)+2K_{ij}h^{ji}+
	\frac{2}{N}\partial_i N^i= \nonumber \\
&&	=\frac{2}{N^2}\partial_0 N+2\partial_i (\frac{N^i}{N})+2K_{ij}h^{ji} \ 
	\nonumber \\
\end{eqnarray}
where we used the fact that 
\begin{eqnarray}
&&	\partial_0 h_{ij}h^{ij}=2N K_{ij}h^{ji}+2\nabla_i N_jh^{ji}=	2NK+2\partial_i N^i+N^i\partial_i h_{kl}h^{kl} \ .  \nonumber \\
\end{eqnarray}
Inserting (\ref{Rde}),(\ref{kindet}) and (\ref{parnulaN}) into (\ref{Sdet})
we obtain 
\begin{eqnarray}\label{Sdecomfin}
&&	S=\frac{1}{\kappa}\int d^nx [\sqrt{-g}K^{ij}\mG_{ijkl}K^{kl}
	+\sqrt{-g}r+2\partial_\mu (N\sqrt{h}n^\mu K)-2\partial_i[\sqrt{h}h^{ij}
	\partial_j N]-\nonumber \\
&&	-\frac{G\sqrt{-g}}{g^2N^2}(\partial_0 g-N^i\partial_i g)^2+\frac{G\sqrt{-g}}{g^2}
	h^{ij}\partial_i g\partial_j g]=\nonumber \\
&&	=\frac{1}{\kappa}\int d^nx [\sqrt{-g}K_{ij}\mG^{ijkl}K_{kl}
	+\sqrt{-g}r
	+\frac{\sqrt{-g}G}{g^2}
	h^{ij}\partial_i g\partial_j g-\nonumber \\
	&&	-4\sqrt{-g}G(\frac{1}{N^2}\partial_0N+\partial_i(\frac{N^i}{N})+K)^2]
\ , \nonumber \\
\end{eqnarray}
where we ignored total derivative terms and where we introduced metric $\mG_{ijkl}$ defines as
\begin{equation}
	\mG^{ijkl}=\frac{1}{2}(h^{ik}h^{jl}+h^{il}h^{jk})-h^{ij}h^{kl} \ 
\end{equation}
that has inverse $\mG_{ijkl}$ 
\begin{equation}
	\mG_{ijkl}=\frac{1}{2}(h_{ik}h_{jl}+h_{il}h_{jk})-\frac{1}{n-2}h_{ij}h_{kl} \ 
\end{equation}
so that
\begin{equation}
	\mG_{ijkl}\mG^{klmn}=\frac{1}{2}(\delta_i^m\delta_j^n+\delta_i^n\delta_j^m) \ .
\end{equation}

Now we are ready to proceed to the canonical formulation of theory. 
From the action (\ref{Sdecomfin}) we derive conjugate momenta 
\begin{eqnarray}\label{defmom}
&&	\pi^{ij}=\frac{\partial \mL}{\partial(\partial_0 h_{ij})}=\frac{\sqrt{-g}}{\kappa N}\mG^{ijkl}K_{kl}-\frac{8}{\kappa}\sqrt{-g}G\tX\frac{1}{2N}h^{ij} \ , \nonumber \\
&&	\pi_N=\frac{\partial \mL}{\partial (\partial_0N)}=-\frac{8}{\kappa}\sqrt{-g}G\frac{1}{N^2}
	\tX \ , \quad 
\tX=	\frac{1}{N^2}\partial_0N+\partial_i(\frac{N^i}{N})+K \ ,
	\nonumber \\	
&&  \pi_i=\frac{\partial \mL}{\partial (\partial_0 N^i)}\approx 0 \ . \nonumber \\	
\end{eqnarray}
Then the Hamiltonian is equal to
\begin{eqnarray}
&&	\mH=\pi^{ij}\partial_0 h_{ij}+\pi_N\partial_0 N-\mL=
	2N\pi^{ij}K_{ij}+2\pi^{ij}\nabla_i N_j+
	\nonumber \\
&&	+\pi_N(\partial_0 N+N^2\partial_i(\frac{N^i}{N}))-
	\pi_NN^2\partial_i (\frac{N^i}{N}) 
	-\mL=\nonumber \\
&&	=\frac{1}{\kappa}\sqrt{-g}K_{ij}\mG^{ijkl}K_{kl}-
	\frac{4}{\kappa}\sqrt{-g}G\tX^2-\sqrt{-g}r-\frac{\sqrt{-g}G}{g^2}h^{ij}\partial_i g\partial_j g
 +N^i\mH_i\equiv \nonumber \\
 &&\equiv \mH_T+N^i\mH_i \ , \nonumber \\
\end{eqnarray}
where we implicitly used integration by parts to introduced $\mH_i$ in the form
\begin{equation}
	\mH_i=-2\nabla_m \pi^{mk}h_{ki}+\partial_i\pi_NN+
	2\pi_N\partial_iN \ . 
\end{equation}
To proceed further we use relation between $\tX$ and $\pi_N$ given on the second line  in (\ref{defmom}) 
 in the definition of $\pi^{ij}$ and we obtain  
\begin{equation}
	\pi^{ij}-\frac{1}{2}\pi_NNh^{ij}=
	\frac{1}{\kappa N}\sqrt{-g}\mG^{ijkl}K_{kl} 
\end{equation}
that allows us to express $K_{ij}$ as 
\begin{equation}
	K_{ij}=\frac{\kappa N}{\sqrt{-g}}\mG_{ijkl}(\pi^{kl}-\frac{1}{2}\pi_NNh^{kl}) \ .
\end{equation}
Using these  results we finally express $\mH_T$ as function of canonical variables 
\begin{eqnarray}
&&	\mH_T=\frac{\kappa}{\sqrt{-g}} N^2
	(\pi^{ij}-\frac{1}{2}\pi_NNh^{ij})
	\mG_{ijkl}(\pi^{kl}-\frac{1}{2}\pi_NN h^{kl})-\nonumber \\
&&	-\frac{\kappa}{16}\frac{1}{\sqrt{-g}G}N^4\pi_N^2
	-\sqrt{-g}r-\frac{\sqrt{-g}G}{g^2\kappa}h^{ij}\partial_i g\partial_j g	=
	\nonumber \\	
&&	\frac{\kappa}{\sqrt{-g}}N^2\left[\pi^{ij}\mG_{ijkl}\pi^{kl}+
	\frac{1}{n-2}\pi_NN\pi-\frac{1}{4}
	\pi_N^2N^2\frac{(n-1)}{(n-2)}\right]-\nonumber \\
&&	-\frac{\kappa}{16\sqrt{-g}G}N^4\pi_N^2-\sqrt{-g}r-\frac{\sqrt{-g}G}{g^2\kappa}
	h^{ij}\partial_i g\partial_j g \ . \nonumber \\
\end{eqnarray}
Then the Hamiltonian with primary constraints $\pi_i\approx 0$ included has the form
\begin{equation}
H=\int d^{n-1}\bx (\mH_T+N^i\mH_i+v^i\pi_i) \ . 
\end{equation}
In the next section we use this form of the Hamiltonian for the study of the stability of constraints. 
\section{Stability of Constraints}\label{third}
In this section we study stability of the primary constraints $\pi_i\approx 0$. 
By definition the constraint is stable if it is preserved during the time evolution of the system
\begin{equation}
\partial_t \pi_i=\pb{\pi_i,H}=
-\mH_i \approx 0
\end{equation}
and we see that the requirement of the preservation of constraint $\pi_i\approx 0$ implies secondary constraints 
\begin{equation}
\mH_i\approx 0 \ . 
\end{equation}
For further purposes we introduce its smeared form defined as
\begin{equation}
\bT_S(\xi)=\int d^{n-1}\bx\xi^i(\bx)\mH_i(\bx)
\end{equation}
and calculate Poisson bracket between $\bT_S(\xi)$  and fundamental fields. Explicitly we get 
\begin{eqnarray}
&&\pb{\bT_S(\xi),h_{ij}}=-\xi^m\partial_m h_{ij}-\partial_i \xi^m
h_{mj}-h_{im}\partial_j \xi^m \ , \nonumber \\
&&\pb{\bT_S(\xi),N}=-\xi^i\partial_i N+\partial_i \xi^i N \ , \nonumber \\
&&\pb{\bT_S(\xi),\pi^{ij}}
=-\partial_m(\xi^m \pi^{ij})+\partial_m\xi^i\pi^{mj}+
\pi^{im}\partial_m\xi^j \ , \nonumber \\
&&\pb{\bT_S(\xi),\pi_N}=	-\partial_i\pi_N\xi^i-2\pi_N\partial_i \xi^i \ , 
\nonumber \\
&&\pb{\bT_S(\xi) h}=-\xi^m\partial_mh-2\partial_i\xi^i h \ ,  \nonumber \\
&&\pb{\bT_S(\xi),g}=-
\xi^m\partial_m g \ , \nonumber \\
&&\pb{\bT_S(\xi),r}=-\xi^m\partial_m r \ . \nonumber \\
\end{eqnarray}
We see that formally $N$ behaves as scalar density of weight $-1$ while $\pi_N$ is scalar density of the weight $+2$. Further, there is an important result that shows that $g=-N^2h$ transforms as scalar when we calculate its Poisson bracket with $\bT_S(\xi)$.

Now collecting all terms together we get that 
\begin{eqnarray}
\pb{\bT_S(\xi),\mH_T}=-\xi^m\partial_m \mH_T \ . 
\nonumber \\
\end{eqnarray}
In other words $\mH_T$ behaves as scalar rather than as scalar density when we calculate its Poisson bracket with $\bT_S(\xi)$. 
 
Now we can study time evolution of the constraint $\mH_i\approx 0$ or its smeared
form. Explicitly the evolution of  the constraint $\bT_S(\xi)$ has the form 
\begin{eqnarray}
&&\partial_t \bT_S(\xi)=\pb{\bT_S(\xi),H}=\int d^{n-1}\by\pb{\bT_S(\xi),\mH_T(\by)}+
\pb{\bT_S(\xi),\bT_S(N^j)}=\nonumber \\
&&=-\int d^{n-1}\by\xi^m\partial_m\mH_T
+
\bT_S((\xi^j\partial_j N^i-N^j\partial_j\xi^i)) \approx -\int d^{n-1}\by \xi^m
\partial_m \mH_T \ , \nonumber \\
\end{eqnarray}
where we used the fact that the  Poisson bracket between 
two smeared forms of the constraints  $\bT_S(\xi^i)$ and $\bT_S(N^i)$ is equal to 
\begin{eqnarray}
\pb{\bT_S(\xi^i),\bT_S(N^j)}
=\bT_S((\xi^j\partial_j N^i-N^j\partial_j\xi^i)) \ . \nonumber \\
\end{eqnarray}
We see that the constraint $\mH_i\approx 0$ is preserved on condition when we introduce 
tertiary constraint 
\begin{equation}
\mH_T(\bx)\approx 0 \ .
\end{equation}
Note that there is  difference between covariant theories of gravity when Hamiltonian constraint emerges as a consequence of the requirement of the preservation of the primary 
constraint $\pi_N\approx 0$ while  in case of the theory with restricted coordinate invariance $\mH_T\approx 0$ emerges as the tertiary constraint when we require preservation of constraints $\mH_i\approx 0$.

As the last step we should study the condition of the preservation of the constraint $\mH_T\approx 0$. As in case of the spatial diffeomorphism constraint we introduce its 
smeared form 
\begin{equation}
\bT_T(X)=\int d^{n-1}\bx X(\bx) \mH_T(\bx) \ . 
\end{equation}
First of all we obtain 
\begin{eqnarray}
&&\pb{\bT_S(\xi^i),\bT_T(X)}=
\int d^{n-1}\bx X (-\xi^m\partial_m \mH)=\nonumber \\
&&\int d^{n-1}\bx\partial_m(\xi^m X)\mH=
\bT_T(\partial_m(\xi^m X)) \  \nonumber \\
\end{eqnarray}
and hence it vanishes on the constraint surface $\mH_T\approx 0$. 

As the next step we will calculate  Poisson bracket between two  smeared forms of Hamiltonian constraints $\bT(X)\approx 0,\bT(Y) \approx 0$,  
where $X(\bx),Y(\bx)$ are arbitrary test functions. Since this is one of  the most important parts of this paper we would like to show these calculations in more details. 
First of all we have
\begin{eqnarray}
&&	\pb{\int d^{n-1}\bx X\frac{\kappa}{4\sqrt{-g}}N^4\pi_N^2(\frac{(n-1)}{(n-2)}+\frac{1}{4G}),\int
		d^{n-1}\by YG\frac{\sqrt{-g}}{g^2\kappa}h^{ij}\partial_ig\partial_jg}+
	\nonumber \\
&&	+\pb{
		\int
		d^{n-1}\bx X\frac{\sqrt{-g}G}{g^2\kappa}h^{ij}\partial_ig\partial_jg	,
		\int d^{n-1}\by Y\frac{\kappa}{4\sqrt{-g}G}N^4\pi_N^2(\frac{(n-1)}{(n-2)}+\frac{1}{4G})}=
	\nonumber \\
&&	=-2\int d^{n-1}\bx (X\partial_i Y-Y\partial_iX)N^5\pi_N\frac{h}{g^2}
	\left(\frac{(n-1)}{(n-2)}+\frac{1}{4G}\right)Gh^{ij}\partial_jg \ .  \nonumber \\
\end{eqnarray}
We further have 
\begin{eqnarray}
&&	\pb{\int d^{n-1}\bx X\frac{\kappa}{\sqrt{-g}}N^2\pi^{ij}\mG_{ijkl}\pi^{kl} , 
		-\int	d^{n-1}\by YG\frac{\sqrt{-g}}{g^2\kappa}h^{ij}\partial_ig\partial_jg}+
	\nonumber \\
&&	+\pb{
		-\int
		d^{n-1}\bx X\frac{\sqrt{-g}G}{g^2\kappa}h^{ij}\partial_ig\partial_jg	,
		\int d^{n-1}\by Y\frac{\kappa}{\sqrt{-g}G}\pi^{ij}\mG_{ijkl}\pi^{kl}}=\nonumber \\
&&	=-4\int d^{n-1}\bx (X\partial_i Y-Y\partial_iX)h^{ij}\partial_j g\frac{1}{g^2(n-2)}GN^4h\pi \  \nonumber \\
\end{eqnarray}
and also
\begin{eqnarray}
&&	\pb{\int d^{n-1}\bx X \frac{\kappa}{(n-2)\sqrt{-g}}\pi_NN^3\pi,
		-\int d^{n-1}\by Y \frac{\sqrt{-g}G}{g^2\kappa}h^{ij}\partial_i g\partial_j g}+
	\nonumber \\
&&	+\pb{-\int  d^{n-1}\bx X\frac{\sqrt{-g}G}{g^2\kappa}h^{ij}\partial_i g\partial_j g,
		\int d^{n-1}\by Y\frac{\kappa}{(n-2)\sqrt{-g}}\pi_NN^3\pi_N}=\nonumber \\
&&	=4\int d^{n-1}\bx (
	X\partial_iY-Y\partial_i X)\frac{N^4\pi}{(n-2)g^2}Gh^{ij}h\partial_j g+
	\nonumber \\
&&	+2\int d^{n-1}\bx (X\partial_iY-Y\partial_iX)\frac{G(n-1)}{(n-2)g^2}N^5\pi_Nhh^{ij}\partial_jg
	\nonumber \\
\end{eqnarray}
using
\begin{equation}
	\pb{\pi^{ij}(\bx),\sqrt{h}(\by)}=-\frac{1}{2}h^{ij}(\bx)\sqrt{h}\delta(\bx-\by) \ . 
\end{equation}
As the next step we further calculate
\begin{eqnarray}
&&	\pb{\int d^{n-1}\bx X\frac{\kappa}{\sqrt{-g}}N^2\pi^{ij}\mG_{ijkl}\pi^{kl} ,-\int d^{n-1}\by 
		Y\frac{1}{\kappa}\sqrt{-g}r}+
	\nonumber \\
&&	+\pb{-\int d^{n-1}\bx X\frac{1}{\kappa}\sqrt{-g}r,\int d^{n-1}\by
		Y\frac{\kappa}{\sqrt{-g}}N^2\pi^{ij}\mG_{ijkl}\pi^{kl}}=
	\nonumber \\
&&=\int d^{n-1}\bx (X\partial_mY-Y\partial_mX)(-2N^2\nabla_n\pi^{mn})  \  
\nonumber \\
\end{eqnarray}
using following formula
\begin{equation}
	\frac{\delta r(\bx)}{\delta h_{ij}(\by)}=-r^{ij}(\bx)\delta(\bx-\by)
	+\nabla^i\nabla^j\delta(\bx-\by)-h^{ij}(\bx)\nabla_k\nabla^k\delta(\bx-\by) \ . 
\end{equation}
and also the fact that $\nabla_m(\sqrt{-g})=\sqrt{h}\nabla_m N$.  Finally we calculate
\begin{eqnarray}
&&	\pb{\int d^{n-1}\bx X\frac{\kappa}{(n-2)\sqrt{-g}}\pi_NN^3\pi,
		-\int d^{n-1}\by Y\frac{1}{\kappa}\sqrt{-g}r}+\nonumber \\
&&	+\pb{-\int d^{n-1}\bx X\frac{1}{\kappa}\sqrt{-g}r,
		\int d^{n-1}\by Y\frac{\kappa}{(n-2)\sqrt{-g}}\pi_NN^3\pi}=\nonumber \\
&&=\int d^{n-1}\bx (X\partial_iY-Y\partial_iX)h^{ij}N^2\pi_N\partial_jN+
\nonumber \\
&&	+\int d^{n-1}\bx (X\partial_iY-Y\partial_iX)h^{ij}N^3\left(\partial_j\pi_N-\frac{1}{2}\frac{\partial_j h}{h}\pi_N\right) \ . 
	\nonumber \\
\end{eqnarray}
Collecting all these calculations  together we obtain 
\begin{eqnarray}
	\pb{\bT_T(X),\bT_T(Y)}=
	\int d\bx (X\partial_i Y-Y\partial_iX)h^{ij}N^2\mH_j=\bT_S((X\partial_iY-Y\partial_iX)h^{ij}N^2) \ . \nonumber \\
\end{eqnarray}
We derived central result of our analysis that show that the theory possesses $n$ first class constraints $\mH_i\approx 0, \mH_T\approx 0$ together with $(n-1)$ first class constraints  $\pi_i\approx 0$. 
In other words the constraint structure is the same as in case of ordinary General Relativity 
with the important difference that the lapse function $N$ is dynamical. In other words the primary constraints $\pi_i\approx 0$ and corresponding gauge fixing functions allow us to eliminate $N^i$ from the theory. Further, $n$ first class constraints $\mH_T\approx 0,\mH_i\approx 0$ can be gauge fixed in principle. These $n$ constraints together with $n$ gauge fixing functions can be explicitly solved for $n$ components of metric $h_{ij}$ and their conjugate momenta leaving $n(n-3)$ phase space independent physical modes that corresponds  to the phase space degrees of freedom for massless graviton in $n$-dimensions. On the other hand there is still dynamical 
field $N$ with corresponding conjugate momentum $\pi_N$. This mode can be further fixed when the theory with restricted diffeomorphism invariance possesses additional gauge symmetry. Such a famous example is Weyl transverse gravity whose canonical analysis is currently under progress. 


{\bf Acknowledgement:}

This work  is supported by the grant “Dualitites and higher order derivatives” (GA23-06498S) from the Czech Science Foundation (GACR).

%



\begin{thebibliography}{20}


\bibitem{Buchmuller:1988wx}
W.~Buchmuller and N.~Dragon,
\emph{``Einstein Gravity From Restricted Coordinate Invariance,''}
Phys. Lett. B \textbf{207} (1988), 292-294
doi:10.1016/0370-2693(88)90577-1







\bibitem{Henneaux:1989zc}
M.~Henneaux and C.~Teitelboim,
\emph{``The Cosmological Constant and General Covariance,''}
Phys. Lett. B \textbf{222} (1989), 195-199
doi:10.1016/0370-2693(89)91251-3



\bibitem{Kuchar:1991xd}
K.~V.~Kuchar,
\emph{``Does an unspecified cosmological constant solve the problem of time in quantum gravity?,''}
Phys. Rev. D \textbf{43} (1991), 3332-3344
doi:10.1103/PhysRevD.43.3332


\bibitem{Unruh:1988in}
W.~G.~Unruh,
\emph{``A Unimodular Theory of Canonical Quantum Gravity,''}
Phys. Rev. D \textbf{40} (1989), 1048
doi:10.1103/PhysRevD.40.1048




\bibitem{Jirousek:2023gzr}
P.~Jirou\v{s}ek,
\emph{``Unimodular Approaches to the Cosmological Constant Problem,''}
Universe \textbf{9} (2023) no.3, 131
doi:10.3390/universe9030131
[arXiv:2301.01662 [gr-qc]].

\bibitem{Alvarez:2023utn}
E.~Alvarez and E.~Velasco-Aja,
\emph{``A Primer on Unimodular Gravity,''}
[arXiv:2301.07641 [gr-qc]].

\bibitem{Carballo-Rubio:2022ofy}
R.~Carballo-Rubio, L.~J.~Garay and G.~Garc\'\i{}a-Moreno,
\emph{``Unimodular gravity vs general relativity: a status report,''}
Class. Quant. Grav. \textbf{39} (2022) no.24, 243001
doi:10.1088/1361-6382/aca386
[arXiv:2207.08499 [gr-qc]].






\bibitem{Garay:2023nco}
L.~J.~Garay and G.~Garc\'\i{}a-Moreno,
\emph{``Embedding Unimodular Gravity in string theory,''}
JHEP \textbf{03} (2023), 027
doi:10.1007/JHEP03(2023)027
[arXiv:2301.03503 [hep-th]].






\bibitem{Kehagias:2022mik}
A.~Kehagias, H.~Partouche and N.~Toumbas,
\emph{``A unimodular-like string effective description,''}
Nucl. Phys. B \textbf{991} (2023), 116196
doi:10.1016/j.nuclphysb.2023.116196
[arXiv:2212.14659 [hep-th]].

\bibitem{Tiwari:2022ctc}
S.~C.~Tiwari,
\emph{``New approach to unimodular relativity,''}
Phys. Scripta \textbf{98} (2023) no.6, 065303
doi:10.1088/1402-4896/acd6c4
[arXiv:2212.13137 [physics.gen-ph]].

\bibitem{Alonso-Serrano:2021uok}
A.~Alonso-Serrano and M.~Li\v{s}ka,
\emph{``Thermodynamics of spacetime and unimodular gravity,''}
Int. J. Geom. Meth. Mod. Phys. \textbf{19} (2022) no.Supp01, 2230002
doi:10.1142/S0219887822300021
[arXiv:2112.06301 [gr-qc]].

\bibitem{Alonso-Serrano:2020pcz}
A.~Alonso-Serrano and M.~Li\v{s}ka,
\emph{``New perspective on thermodynamics of spacetime: The emergence of unimodular gravity and the equivalence of entropies,''}
Phys. Rev. D \textbf{102} (2020) no.10, 104056
doi:10.1103/PhysRevD.102.104056
[arXiv:2008.04805 [gr-qc]].

\bibitem{Nojiri:2015sfd}
S.~Nojiri, S.~D.~Odintsov and V.~K.~Oikonomou,
\emph{``Unimodular $F(R)$ Gravity,''}
JCAP \textbf{05} (2016), 046
doi:10.1088/1475-7516/2016/05/046
[arXiv:1512.07223 [gr-qc]].










\bibitem{Karataeva:2022mll}
I.~Y.~Karataeva and S.~L.~Lyakhovich,
\emph{``Gauge symmetry of unimodular gravity in Hamiltonian formalism,''}
Phys. Rev. D \textbf{105} (2022) no.12, 124006
doi:10.1103/PhysRevD.105.124006
[arXiv:2203.06620 [hep-th]].


\bibitem{Bufalo:2017tms}
R.~Bufalo and M.~Oksanen,
\emph{``Canonical structure and extra mode of generalized unimodular gravity,''}
Phys. Rev. D \textbf{97} (2018) no.4, 044014
doi:10.1103/PhysRevD.97.044014
[arXiv:1712.09535 [hep-th]].

\bibitem{Bufalo:2015wda}
R.~Bufalo, M.~Oksanen and A.~Tureanu,
\emph{``How unimodular gravity theories 
	differ from general relativity at quantum level,''}
Eur. Phys. J. C \textbf{75} (2015) no.10, 477
doi:10.1140/epjc/s10052-015-3683-3
[arXiv:1505.04978 [hep-th]].

\bibitem{Kluson:2014esa}
J.~Kluson,
\emph{``Canonical Analysis of Unimodular Gravity,''}
Phys. Rev. D \textbf{91} (2015) no.6, 064058
doi:10.1103/PhysRevD.91.064058
[arXiv:1409.8014 [hep-th]].






\bibitem{Alvarez:2023utn}
E.~Alvarez and E.~Velasco-Aja,
\emph{``A Primer on Unimodular Gravity,''}
[arXiv:2301.07641 [gr-qc]].






\bibitem{Alvarez:2006uu}
E.~Alvarez, D.~Blas, J.~Garriga and E.~Verdaguer,
\emph{``Transverse Fierz-Pauli symmetry,''}
Nucl. Phys. B \textbf{756} (2006), 148-170
doi:10.1016/j.nuclphysb.2006.08.003
[arXiv:hep-th/0606019 [hep-th]].

\bibitem{Bonifacio:2015rea}
J.~Bonifacio, P.~G.~Ferreira and K.~Hinterbichler,
\emph{``Transverse diffeomorphism and Weyl invariant massive spin 2: Linear theory,''}
Phys. Rev. D \textbf{91} (2015), 125008
doi:10.1103/PhysRevD.91.125008
[arXiv:1501.03159 [hep-th]].



\bibitem{Oda:2016vui}
I.~Oda,
\emph{``Cosmology in Weyl Transverse Gravity,''}
Mod. Phys. Lett. A \textbf{31} (2016) no.39, 1650218
doi:10.1142/S0217732316502187
[arXiv:1609.00407 [gr-qc]].






\bibitem{Oda:2016psn}
I.~Oda,
\emph{``Classical Weyl Transverse Gravity,''}
Eur. Phys. J. C \textbf{77} (2017) no.5, 284
doi:10.1140/epjc/s10052-017-4843-4
[arXiv:1610.05441 [hep-th]].




\bibitem{Alonso-Serrano:2022pif}
A.~Alonso-Serrano, L.~J.~Garay and M.~Li\v{s}ka,
\emph{``Noether charge formalism for Weyl invariant theories of gravity,''}
Phys. Rev. D \textbf{106} (2022) no.6, 064024
doi:10.1103/PhysRevD.106.064024
[arXiv:2206.08746 [gr-qc]].

\bibitem{Alonso-Serrano:2022rzj}
A.~Alonso-Serrano, L.~J.~Garay and M.~Li\v{s}ka,
\emph{``Noether charge formalism for Weyl transverse gravity,''}
Class. Quant. Grav. \textbf{40} (2023) no.2, 025012
doi:10.1088/1361-6382/acace3
[arXiv:2204.08245 [gr-qc]].


\bibitem{Gourgoulhon:2007ue}
E.~Gourgoulhon,
\emph{``3+1 formalism and bases of numerical relativity,''}
[arXiv:gr-qc/0703035 [gr-qc]].

\bibitem{Arnowitt:1962hi}
R.~L.~Arnowitt, S.~Deser and C.~W.~Misner,
\emph{``The Dynamics of general relativity,''}
Gen. Rel. Grav. \textbf{40} (2008), 1997-2027
doi:10.1007/s10714-008-0661-1
[arXiv:gr-qc/0405109 [gr-qc]].


\end{thebibliography}
\end{document}